\documentclass[11pt,a4paper,showpacs,showkeys,,superscriptaddress]{article}

\usepackage{epsfig}
\usepackage{amsmath,amssymb}
\usepackage{graphicx}
\usepackage{xcolor}
\usepackage{subfigure}

\usepackage{latexsym}
\usepackage{hyperref}
\hypersetup{colorlinks,%
  citecolor=blue,%
  linkcolor=red,%
  pdftex}

\begin{document}

\baselineskip 6mm
\renewcommand{\thefootnote}{\fnsymbol{footnote}}

\newcommand{\nc}{\newcommand}
\newcommand{\rnc}{\renewcommand}



\newcommand{\tcb}{\textcolor{blue}}
\newcommand{\tcr}{\textcolor{red}}
\newcommand{\tcg}{\textcolor{green}}


\def\beq{\begin{equation}}
\def\eeq{\end{equation}}
\def\ba{\begin{array}}
\def\ea{\end{array}}
\def\bea{\begin{eqnarray}}
\def\eea{\end{eqnarray}}
\def\nn{\nonumber}


\def\CMP{Commun. Math. Phys.~}
\def\JHEP{JHEP~}
\def\Pre{Preprint}
\def\PRL{Phys. Rev. Lett.~}
\def\PR {Phys. Rev.~}
\def\CQG {Class. Quant. Grav.~}
\def\PL {Phys. Lett.~}
\def\NP {Nucl. Phys.~}

\def\G{\Gamma}

\def\S{{\bf S}}
\def\C{{\bf C}}
\def\Z{{\bf Z}}
\def\R{{\bf R}}
\def\N{{\bf N}}
\def\M{{\bf M}}
\def\P{{\bf P}}
\def\bm{{\bf m}}
\def\bn{{\bf n}}

\def\CA{{\cal A}}
\def\CB{{\cal B}}
\def\CC{{\cal C}}
\def\CD{{\cal D}}
\def\CE{{\cal E}}
\def\CF{{\cal F}}
\def\CH{{\cal H}}
\def\CM{{\cal M}}
\def\CG{{\cal G}}
\def\CI{{\cal I}}
\def\CJ{{\cal J}}
\def\CL{{\cal L}}
\def\CK{{\cal K}}
\def\CN{{\cal N}}
\def\CO{{\cal O}}
\def\CP{{\cal P}}
\def\CQ{{\cal Q}}
\def\CR{{\cal R}}
\def\CS{{\cal S}}
\def\CT{{\cal T}}
\def\CU{{\cal U}}
\def\CV{{\cal V}}
\def\CW{{\cal W}}
\def\CX{{\cal X}}
\def\CY{{\cal Y}}
\def\CZ{{\cal Z}}

\def\We{{W_{\mbox{eff}}}}


\newcommand{\Lie}{\pounds}

\newcommand{\p}{\partial}
\newcommand{\bp}{\bar{\partial}}

\newcommand{\half}{\frac{1}{2}}

\newcommand{\bfalpha}{{\mbox{\boldmath $\alpha$}}}
\newcommand{\bfbeta}{{\mbox{\boldmath $\beta$}}}
\newcommand{\bfgamma}{{\mbox{\boldmath $\gamma$}}}
\newcommand{\bfmu}{{\mbox{\boldmath $\mu$}}}
\newcommand{\bfpi}{{\mbox{\boldmath $\pi$}}}
\newcommand{\bfvarpi}{{\mbox{\boldmath $\varpi$}}}
\newcommand{\bftau}{{\mbox{\boldmath $\tau$}}}
\newcommand{\bfeta}{{\mbox{\boldmath $\eta$}}}
\newcommand{\bfxi}{{\mbox{\boldmath $\xi$}}}
\newcommand{\bfkappa}{{\mbox{\boldmath $\kappa$}}}
\newcommand{\bfepsilon}{{\mbox{\boldmath $\epsilon$}}}
\newcommand{\bfTheta}{{\mbox{\boldmath $\Theta$}}}

\newcommand{\bz}{{\bar{z}}}

\newcommand{\dalpha}{\dot{\alpha}}
\newcommand{\dbeta}{\dot{\beta}}
\newcommand{\blambda}{\bar{\lambda}}
\newcommand{\btheta}{{\bar{\theta}}}
\newcommand{\bsigma}{{{\bar{\sigma}}}}
\newcommand{\bepsilon}{{\bar{\epsilon}}}
\newcommand{\bpsi}{{\bar{\psi}}}


\def\ct{\cite}
\def\la{\label}
\def\eq#1{(\ref{#1})}


\def\a{\alpha}
\def\b{\beta}
\def\g{\gamma}
\def\G{\Gamma}
\def\d{\delta}
\def\D{\Delta}
\def\ep{\epsilon}
\def\e{\eta}
\def\ph{\phi}
\def\Ph{\Phi}
\def\ps{\psi}
\def\Ps{\Psi}
\def\k{\kappa}
\def\l{\lambda}
\def\L{\Lambda}
\def\m{\mu}
\def\n{\nu}
\def\th{\theta}
\def\Th{\Theta}
\def\r{\rho}
\def\s{\sigma}
\def\S{\Sigma}
\def\ta{\tau}
\def\o{\omega}
\def\O{\Omega}
\def\pr{\prime}
\def\f{\varphi}


\def\half{\frac{1}{2}}

\def\goto{\rightarrow}

\def\na{\nabla}
\def\grad{\nabla}
\def\curl{\nabla\times}
\def\div{\nabla\cdot}
\def\pa{\partial}

\def\bra{\left\langle}
\def\ket{\right\rangle}
\def\lb{\left[}
\def\lc{\left\{}
\def\ls{\left(}
\def\lp{\left.}
\def\rp{\right.}
\def\rb{\right]}
\def\rc{\right\}}
\def\rs{\right)}
\def\cl{\mathcal{l}}

\def\vac#1{\mid #1 \rangle}

\def\td#1{\tilde{#1}}
\def\check{ \maltese {\bf Check!}}


\def\Tr{{\rm Tr}\,}
\def\det{{\rm det}\,}


\def\bc#1{\nnindent {\bf $\bullet$ #1} \\ }
\def\ch {$<Check!>$ }
\def\ss {\vspace{1.5cm}}

\begin{titlepage}
%
%
%
%
%
%
%
%
\begin{center}
{\Large \bf Holography without counter terms}
%
\vskip 1. cm
  { Byoungjoon Ahn\footnote{e-mail : bjahn@yonsei.ac.kr}, Seungjoon Hyun\footnote{e-mail : sjhyun@yonsei.ac.kr}, Kyung Kiu Kim\footnote{e-mail : kimkyungkiu@gmail.com}, Sang-A Park\footnote{e-mail : sangapark@yonsei.ac.kr},
  Sang-Heon Yi\footnote{e-mail : shyi@yonsei.ac.kr}
 }
\vskip 0.5cm
{\it Department of Physics, College of Science, Yonsei University, Seoul 120-749, Korea}\\
\end{center}
\thispagestyle{empty}
\vskip1.5cm
%
%
\centerline{\bf ABSTRACT} \vskip 4mm
 \vspace{1cm}
\noindent
By considering the  behavior of  the reduced action under the scaling transformation,
we present a unified derivation of the Smarr-like relation for asymptotically anti-de-Sitter planar black holes. This novel Smarr-like relation leads to useful information in the condensed matter systems through the AdS/CMT correspondence.
By using our results, we  provide an  efficient way to obtain the holographically  renormalized on-shell action without the information on the explicit forms of counter terms.
We find the complete consistency of our results with those in various models discussed in the recent literatures and obtain new implications.

\vspace{2cm}
%
%
\end{titlepage}
\setcounter{footnote}{0}
%
%
%
%

\section{Introduction}
The anti-de Sitter/condensed matter theory(AdS/CMT) correspondence  becomes an interesting and actively developing subject after its introduction, which is realized as a concrete application of the AdS/CFT correspondence~\cite{Maldacena:1997re}. See \cite{Hartnoll:2009sz,Horowitz:2010gk,Herzog:2009xv,Sachdev:2010ch} for reviews on the AdS/CMT correspondence.
Through the correspondence, there have been many interesting studies on phenomena in condensed matter theory, specifically such as DC/AC conductivities for lattice models and models with momentum relaxation, metal insulator transitions, Hall angle, Nernst signal, incoherent metal transition and so on. See \cite{Horowitz:2012ky,Donos:2014uba,Donos:2014cya,Vegh:2013sk,Donos:2012js,Blake:2014yla,Kim:2014bza,Kim:2015wba} and their related works.

Basically, this new approach to  the strong coupling behavior of  field theory systems goes beyond our perturbative understanding of the conventional field theory approach  and so provides us a new perspective to the understanding of strong coupling physics, at least qualitatively. One of the advantages in this approach is that the hidden  universality in the condensed matter system  may be revealed in a rather simple way.  According to the AdS/CFT correspondence, the universal  information on the classical bulk gravity  implies such a  universality in the  dual quantum field theory system. For instance, the field theory systems,  apparently disparate from each other, would  have a universal thermodynamic behavior  since the dual black hole thermodynamics has been known to be  universal, irrespectively of the types of black holes or the gravity Lagrangian.

In the context of the AdS/CMT correspondence, it would be very interesting to understand the universal behaviors of various holographic models.  One of such universal features is, of course, the first law of black hole thermodynamics,  which has already been explored extensively in various cases.  Recently, the Smarr relation or its extension are emphasized as a kind of universal property of planar black holes~\cite{Hyun:2015tia,Ahn:2015uza}. This universal property in its simplest context comes from the generic existence of the scaling invariance in the reduced action of the model with some spherical/planar symmetry~\cite{Banados:2005hm, Hyun:2015tia}. In this case, the charge  associated with the scaling invariance  can be connected with the entropy at the horizon and 
the mass  and the charge at
 the asymptotic infinity and thus could be used to establish the Smarr relation.  Even more striking case is the one which lacks the spherical/planar symmetry, and thus whose reduced action does not have the scaling invariance. It has been found in \cite{Ahn:2015uza} that, even in those cases, one can still consider the scaling transformation on the reduced action and extract the Smarr-like relation. In this case, the associated charge-like quantity coined as the charge function,  
from the scaling  transformation in the reduced action 
 can, still, be related to the entropy at the horizon and  the mass at the infinity.  We would like to emphasize that our methods utilizing the behavior of the reduced action under the scaling transformation are  rather generic, not specific to the models considered in the main text, and could be extended to more general cases.

  Unlike the original Smarr relation in the asymptotically flat spacertime, which was derived by a dimensional analysis on physical quantities  of black holes~\cite{Smarr:1972kt}, the existence of cosmological constant defies such a derivation. This disparity was explicitly stated  in Ref.~\cite{Gibbons:2004ai}  as  ``the failure of the Smarr-Gibbs-Duhem relation in the presence of the cosmological constant''.   Furthermore,  the divergence in the asymptotic AdS spacetime requires careful treatment which is now well-established as the holographic renormalization procedure. Because of this difference, the quantum statistical relation rather than the Smarr-Gibbs-Duhem relation is focused on in the asymptotically AdS black holes~\cite{Gibbons:2004ai,Papadimitriou:2005ii}. 
 
On the other hand, it has been known that  one can obtain the Smarr(-like) relation for Banados-Teitelboim-Zanelli black holes~\cite{Banados:1992wn} and specific asymptotically AdS planar black holes~\cite{Horowitz:2010gk}. In this paper, we provide a generic derivation of the Smarr-like relation for the asymptotically AdS planar black holes, purely in the bulk gravity context.   In the context of the AdS/CMT correspondence, the bulk Smarr-like relation for the planar black holes should be matched with the  thermodynamic relation  for boundary system. This match would be the basic check point for a consistent AdS/CMT model construction. For instance, see~\cite{Gauntlett:2009dn,Gauntlett:2009bh,Sonner:2010yx} for such a task in some holographic  models and see also~\cite{Tian:2014goa} for some attempts to understand the Smarr-like relation  from another view point.

The organization  and summary of the paper are as follows. In section 2, we  derive Smarr-like relations by using a scaling transformation  in  a unified way, which includes  various examples obtained, case by case,  through  other methods. We introduce a specific function coined as the charge function to present our Smarr-like relation.  Then, we  propose our Smarr-like relation combined with the background subtraction method  as an efficient formula to calculate the finite renormalized on-shell action. In particular,  this may  be very useful when the analytic solution of the given model is not available.
 In section 3,  we consider various holographic models in four dimensions and  check that our general formula is   consistent with known results.  It turns out that the counter term structure of the holographic renormalization is veiled in the divergences of the charge function in our procedure. However, it does not affect our Smarr-like relations, since the divergences should cancel by construction.  By adopting the background subtraction method, we obtain the  finite on-shell actions for extended Einstein-Maxwell-dilaton models. 
Our method can be easily extended  beyond AdS geometries as was done in Ref.~\cite{Hyun:2015tia}, though they are not discussed explicitly in this paper. Section 4 is devoted to conclusions and the future direction. In Appendix A, we generalize the formulas   when  some higher form fields exist in arbitrary dimensions. In Appendix B,  we provide some details on how to obtain the renormalized on-shell action and the thermodynamic relation using the  conventional holographic renormalization and the counter terms.

\section{The reduced action and the scaling transformation}

In this section we give the details of our method to obtain the Smarr-like relation of various planar black hole solutions in \cite{Banados:2005hm,Hyun:2015tia,Ahn:2015uza}. We consider the generic Einstein-Maxwell-scalar theory  with the action:
\begin{align}  \label{origaction}
S = \frac{1}{16 \pi  G} \int d^D x \sqrt{-g} \left( R + (D-1)(D-2) -\frac{1}{4}Z(\phi)\mathcal{F}^{\mu\nu}\mathcal{F}_{\mu\nu}-\frac{1}{2}(\partial \phi)^2 -V(\phi) \right),
\end{align}
where the curvature radius of the asymptotic AdS space is taken as the unity. Despite the fact that we are focusing on the case of a single gauge and scalar fields, it is straightforward to extend to the case of arbitrary number of gauge and scalar fields. One may also note that, in fact, our method  can be extended straightforwardly to other asymptotic geometries such as the Lifshitz one~\cite{Hyun:2015tia}. 
Depending on the symmetry of the full solution, the reduced action may or may not have the scaling invariance. On the planar black holes with full planar   symmetry, the reduced action is invariant under rescaling, while on the geometry without full planar invariance, for example, due to the  momentum along the plane the rescaling symmetry is broken. In the first subsection, we review the reduced action formalism and the emergence of  the scaling symmetry for the system with planar symmetry. In the second one, we consider the case with broken scaling symmetry and explain that the scaling argument is still relevant and useful  to reveal the Smarr-like relation.  

 We would like to emphasize that our derivation of the Smarr-like relation is performed  completely in the  bulk gravity and does not utilize the thermodynamic relation at all. In the last part of this section, as an application of the Smarr-like relation, we show how  to obtain  the thermodynamic pressure of the dual field theory system.  According to the AdS/CFT dictionary, the bulk  black hole configuration can be identified with the boundary thermal state. And then the black hole parameters are identified with the thermodynamic quantities of the dual field theory\footnote{We identify the mass of the black hole, the bulk chemical potential, the charge of black hole, the Hawking temperature and the entropy of the black hole with the energy, the chemical potential, the charge density, the temperature and the entropy in the dual system, respectively.}. Now,  let us assume that the AdS/CMT model is consistently constructed and that the boundary thermodynamic relation holds,  for instance,  in the form of   
${\cal P} + {\Large \text{$\epsilon$}} = \mu  Q  + T S$.
By adopting a specific method to obtain the mass of planar black holes, we show that the pressure expression can be determined from the Smarr-like relation combined with  the  above thermodynamic relation. In our case, the pressure is nothing but  the finite renormalized  on-shell action through the AdS/CFT correspondence.

\subsection{The planar symmetry vs. the scaling symmetry}
In this subsection, we  consider the case with full planar symmetry and thus take the ansatz for metric, gauge and scalar fields of planar AdS black holes as
\begin{align}\label{ansatz}
ds^2 &= - r^{-2(D-2)} e^{2A(r)}f(r) dt^2 + \frac{dr^2}{r^2 f(r)} + r^2d{\bf x}^2_{D-2}\,,\\
\CA&= r^{-(D-2)}a(r)dt\,,\qquad\phi = \phi(r)\,,
\end{align}
where ${\bf x}=(x^{1},x^{2},\cdots, x^{(D-2)})$ denote the spatial coordinates other than the radial one. 
 Under the ansatz, the reduced action of our model can be written as
\begin{equation} \label{reduced action}
I_{red}[\Psi] = \frac{1}{16\pi G} \int dt d{\bf x}\, \int dr \,L_{red}(r, \Psi )\,,
\end{equation}
where all the fields, $f, A, \phi, a$,  collectively denoted as $\Psi$,  depend only on the $r$-coordinate.
The associated reduced Lagrangian is given by 
\begin{align}
L_{red}=\frac{e^{A}}{r}\left[-(D-2)\Big(rf'+(D-1)(f-1) \Big) + \frac{1}{2}e^{-2A}Z(\phi) \left( r a' -(D-2)a \right)^2 -\frac{f}{2}r^2 \phi'{}^2 -V(\phi) \right]\,,
\end{align}
where  ${}'$ denotes the derivative with respect to the coordinate $r$ and irrelevant  total derivative terms have been omitted.
Under a generic variation, this reduced action transforms as
\begin{equation} \label{GenVar}
\delta I_{red}[\Psi] =  \frac{1}{16\pi G}\int dt d{\bf x}\, \int dr  \Big[ \CE_{\Psi}\delta \Psi + \partial_r\Theta(\Psi\,;\,\delta \Psi) \Big]\,,
\end{equation}
where the equations of motions(EOM) of the reduced action is given by $\CE_{\Psi}=0$. We assume that these EOM are equivalent with the full EOM from the original action given in Eq.~(\ref{origaction}) for our ansatz.
\\
\\
\underline{\it The scaling symmetry and its associate charge}

It turns out that the reduced Lagrangian has an accidental symmetry which has nothing to do with the diffeomorphism of the original Lagrangian density. Namely, the reduced action is invariant under the scaling transformation, $r \to r e^\sigma$, if all the fields, $f, A, \phi, a$,  collectively denoted as $\Psi$, in the reduced action transform as scalar fields under this transformation. Accordingly, the field variations under the infinitesimal scaling transformation, are given by
\begin{align}
\delta_\sigma \Psi(r) =-\sigma~ r \Psi'(r)\,,
\end{align}
and  the invariance of the  reduced action up to total derivatives is warranted  in the form of
\begin{equation}
 \label{SymVar}
\delta_{\sigma} I_{red}=\frac{ 1}{16\pi G  }\int dt d{\bf x}\int dr ~  \, \partial_r S\,, \qquad S=- \sigma rL_{red}(r, \Psi)\,,
\end{equation}
which could be understood as the consequence of the measure change under the scaling transformtion.
Here, we would like to emphasize that the scaling symmetry is off-shell in the reduced action level, completely independent of the specific solutions, which tells us the possibility to use this formalism in the wide class of solutions including the asymptotically non-AdS spacetime. Furthermore, it would be useful to note that the specific solution does not need to respect this off-shell scaling symmetry. 

By applying the Noether theorem on the scaling symmetry, we find the associated  Noether current, or  nothing but charge in this case, which is conserved along the radial direction as\footnote{As is done  in  the usual Noether charge method, we set  $\sigma=1$ to define the charge in the following.}   
%
%
%
%
%
\begin{align} \label{charge}
c 
&\equiv
\frac{1}{16\pi G }
\Big(  \Th    - S\Big) 
= 
\frac{1}{16\pi G }
\left(\frac{\partial L_{red}}{\partial \Psi'(r)}  \left( - r \Psi'(r) \right)+ r L_{red}\right)\, .
\end{align}
By using the EOM, the conserved charge (density) $c$ can be expressed as
\begin{equation} \label{}
\label{charge function}
c=
\frac{1}{16\pi G}
\left\{e^{A(r)} \Big(  (D-2) r f'(r)+r^2 f(r) \phi '(r)^2\Big)-(D-2) r^{D-1}e^{-A(r)}Z(\phi)  \mathit{a}(r)\Big(r^{2-D} a(r)\Big)'\right\}\,.
\end{equation}
Interestingly enough, this radially conserved charge turns out to be related with various physical conserved quantities. In particular, it is related, in a specific gauge, to the black hole entropy  at the horizon  and the mass and electric charge of the system at the  asymptotic infinity. This gives the great opportunity to relate those physical quantities by using the fact that our charge $c$ remains the same along the radial coordinate $r$. In the various planar black holes studied in \cite{Hyun:2015tia,Ahn:2015uza,Banados:2005hm}, it gives us the Smarr relation among the black hole entropy, the mass and the electric charge.
\\
\\
\underline{\it Comparison at the horizon}

The above ansatz of metric and matter fields admits a stationary Killing vector, $\xi_{T} = \frac{\p}{\p t} $, which becomes null at the horizon and   is related to the horizon entropy {\it a la} Wald~\cite{Wald:1993nt,Iyer:1994ys}.
The location of the horizon is determined  by the condition $f(r_H)=0$, where $r_{H}$ denotes the position of the horizon. One can  identify the   temperature of these black holes as
\begin{equation} \label{}
  T_{H} =  \frac{1}{4\pi}r_H^{-(D-3)}e^{A(r_{H})}f'(r_{H})\,,
\end{equation}
while the entropy (density) of the planar black holes in the Einstein gravity is generically given by the area law
\begin{equation} \label{}
{\small \text{$ {\cal  S}$} } =  \frac{r^{D-2}_{H}}{4G}\,.
\end{equation}
Now, let us choose the gauge of  the $U(1)$ gauge field $\CA_{\mu}$ in such a way that it vanishes at the horizon {\it i.e.} $a(r_H)=0$. Then, the radially conserved charge in Eq.~(\ref{charge function}) becomes, at the horizon, 
\begin{align}
c= \frac{(D-2)}{16\pi G} r_H e^{A(r_{H})}f'(r_{H})\,.
\end{align} 
Therefore, one can immediately identify the charge $c$ at the horizon as the product of the black hole temperature and its entropy as
\begin{equation} \label{}
c(r_H)= (D-2)T_H\mathcal{S} \,.
\end{equation}
\\
\\
\underline{\it Comparison at the asymptotic boundary}

The relevant physical quantities of the non-rotating black holes defined at the  asymptotic infinity are the mass (density), which is the conserved charge for the stationary Killing vector $\xi_{T} = \frac{\p}{\p t} $, and electric charge (density) for the $U(1)$ gauge symmetry. 
The electric charge (density) $\mathcal Q$ of the planar black hole can be straightforwardly obtained as a Noether charge for $U(1)$ symmetry or alternatively as the integral of field strength for the gauge field ${\cal A}_{\mu}(r)$. In our setup, it is given by 
\begin{equation} \label{}
\mathcal Q=
\sqrt{-g}Z(\phi)\mathcal{F}^{tr} \,,
\end{equation}
which can be shown to be conserved along the radial direction by using EOM. It is slightly involved  to obtain the finite mass expression of the gravitational system. Here, we briefly present the mass formula obtained in 
\cite{Hyun:2015tia} in the context of the quasi-local ADT formalism  by using the `on-shell' scaling path  in the parameter path in the solution space (See also~\cite{Horowitz:1999jd,Banados:2005hm}). The infinitesimal form of the mass expression is given by
\begin{align}   \label{TVmass}
 \delta M_{ADT} &= \frac{1}{16\pi G}\Big[ -(D-2) e^A\delta f  -  a r^{-(D-2)} \,\delta \mathcal{Q} -r e^A f  \phi'   \, \delta\phi    \Big]_{r\rightarrow\infty} \,,
\end{align}
where $\delta$, in this particular case, denotes the  parameter variation along the `on-shell' scaling path. We present the final  finite mass expression of planar black holes:
\begin{align}\label{MassBulk}  
(D-1) M = \frac{1}{16\pi G}~ e^{A(r)} \Big[  (D-2) r f'(r)+r^2 f(r) \phi '(r)^2\Big]\Big|_{r\rightarrow\infty}\,.
\end{align}
If  we consider the asymptotic forms of metric and matter fields for asymptotically AdS space, then they are given by 
\begin{align}   \label{asymptotic}
f (r) &= 1+ \cdots -\frac{m}{r^{D-1}}+\cdots \,, \qquad \quad\quad\qquad  \qquad e^{A(r)} = r^{D-1} \Big[1+\cdots\Big]\,, \\
\varphi (r) &- \varphi_{\infty} = \frac{\varphi_{J}}{r^{D-1-\Delta_{\varphi}}}+ \cdots+\frac{\varphi_{O}}{r^{\Delta_{\varphi}}}+ \cdots \,, \qquad~ a(r) = r^{D-2}\Big[16\pi G \mu +\cdots- \frac{\mathcal Q}{r^{\Delta_{A}}} + \cdots\Big]\,, \nn
\end{align}
where $\varphi_{\infty}$ denotes the minimum of the generic scalar potential $V(\phi)$ and $\mu$ denotes the chemical potential conjugate to the electric charge $\mathcal Q$.  
All the examples presented in the next section, the finite ADT mass is simply given by
\begin{align} 
  M = \frac{(D-2)m}{16\pi G}  \,.
\end{align}
On the other hand, 
the expression of charge $c$ at the asymptotic infinity with the above asymptotic behavior of fields is given by
\begin{align}
 c=  \frac{1}{16\pi G}~ e^{A(r)} \Big[  (D-2) r f'(r)+r^2 f(r) \phi '(r)^2\Big]\Big|_{r\rightarrow\infty}-(D-2)  \mu \mathcal Q\,.
\end{align}
This clearly shows that the relation between the charge $c$ and the mass $M$ and the $U(1)$ charge as  
\begin{align}
c = (D-1)M - (D-2) \mu \mathcal{Q} \,.
\end{align}  

Therefore by using the conservation of the charge $c$ along the radial direction we obtain the Smarr relation as
\begin{align}
\frac{D-1}{D-2} M  = \mu \mathcal Q + T_H \mathcal S
\end{align}

\subsection{The scaling transformation and the Smarr relation}
The emergence of scaling symmetry on the gravitational system with planar symmetry and its deep connection with the existence of the Smarr relation is quite remarkable. One may ask whether it could be extended to the gravitational system without the full planar symmetry. This would be a relevant question as, in many interesting models in AdS/CMT correspondence, the planar symmetry is often broken by the boundary condition of the scalar and gauge fields to realize some condensates in dual condensed matter system. 
Surprisingly, we find that one can still use and extend the scaling argument to obtain the Smarr-like relations. 
\\
\\
\underline{\it The reduced action with broken scaling symmetry}
 
In order to describe the reduced action with broken scaling symmetry, it is useful to introduce the weight under the scaling transformation. All the fields, $f, A, \phi, a$, behave as scalar fields under the scaling transformation, $\delta_\sigma \Psi=-r\Psi'$, and thus have the weight $\omega=0$. Then, as we have seen in the previous section, in order for the reduced action to be invariant under the scaling transformation, the Lagrangian should have the weight $\omega=-1$ as
\[
\delta_{\sigma}L = -L-rL'=- (rL)'\,.
\]
On the other hand, if  asymptotic values of the scalar and gauge fields are non-trivial, the reduced action would contain pieces with different weights under the scaling transformation. Specifically, one may decompose the reduced Lagrangian with respect to the weight $\omega$ as
\[
L=\sum_{\omega}L_{[\omega]}\,,
\]
with
\[
\delta_{\sigma}L_{[\omega]} = \omega L_{[\omega]}-rL_{[\omega]}'\,.
\]

 For instance, we may add a massless free scalar field of the configuration as 
\begin{equation} \label{axion}
\varphi = \beta x^{1}\,, \qquad \beta=const.,
\end{equation}
in order to realize a relaxation of the planar symmetry in the dual theory.   In this case, the lack of the planar symmetry breaks the scaling symmetry. Indeed, the additional kinetic term of the scalar field becomes 
\begin{align}
- \sqrt{-g} \frac{1}{2} ( \partial \varphi )^2=- \frac{1}{2} \frac{e^A}{r^3} \beta^2 \nonumber
\end{align}
  in the reduced Lagrangian, which  has weight $\omega=-3$. Furthermore  we may allow  the  various fields  to depend on the spatial coordinates  in such a way that their explicit coordinate dependence does not appear  in the reduced Lagrangian.
\\
\\
\underline{\it The Smarr-like relation with broken scaling symmetry}

In general
 the variation of the Lagrangian under the scaling transformation becomes
\[
\delta_\sigma L= \sum_{\omega} (\omega L_{[\omega]}-rL_{[\omega]}')=- (rL)'+\sum_{\omega} (\omega+1)L_{[\omega]}\,,
\]
and therefore  the variation of the reduced action under the scaling transformation can be expressed as 
\begin{align}\label{ver1}
\delta_{\sigma} I_{red}&=\frac{1}{16\pi G}\int dt d{\bf x}\int dr  \, \Big( \partial_r S + \sum_{\omega}(\omega+1)L_{[\omega]}\Big)\,,
\qquad S=-rL
(r, \Psi)\,.
\end{align}
On the other hand, by using the formula for the variation of the action under the generic field variation, the scaling transformation of the reduced action can also be expressed as
\begin{align}\label{ver2}
\delta_{\sigma} I_{red} =  \frac{1}{16\pi G}\int dt d{\bf x}\int dr \,  \Big( \CE_{\Psi}\delta_\sigma \Psi
+ \partial_r \Theta(\Psi\,;\,\delta_\sigma \Psi) \Big) \,.
\end{align}

By identifying these two expressions on the variation of the reduced action, (\ref{ver1}) and (\ref{ver2}),  one can verify that the
current, $c(r)$, defined in (\ref{charge}), satisfies
\begin{align} \label{partial charge}
 c'(r) = \frac{1}{16\pi G}\sum_{\omega}(\omega+1)L_{[\omega]}\Big|_{on-shell} \,,
\end{align}
where we have used the on-shell condition $\CE_{\Psi}=0$. Surely, the non-vanishing right-hand side tells us that the scaling symmetry of the reduced action is broken and now the current is no longer conserved along the radial direction.  Nevertheless, 
we can still introduce $c(r)$ as the  charge function of the radial coordinate $r$ and  integrate (\ref{partial charge}) to obtain\footnote{An alternative derivation of the Smarr-like relation for the system without the full planar symmetry was given in~\cite{ Ahn:2015uza}. Both descriptions are consistent and give the same results.}  
\begin{equation} \label{ChargeFunc}
c(r) - c(r_{H})= \frac{1}{16\pi G} \int^{r}_{r_{H}} dr  \sum_{\omega}(\omega+1)L_{[\omega]}\Big|_{on-shell} \,,
\end{equation}
which is one of our main results and  would give us a useful information in the holographic models. One nice aspect of the Smarr-like relation is that it gives a very useful criterion to check the consistency of numerical results, when it is not easy to obtain analytic solutions.

Our current/charge function is reminiscent of the axial current in the chiral gauge theory,  which is broken explicitly whenever the fermion field has non-vanishing mass and yet is still useful despite its non-conservation.
The above relation, which we call the Smarr-like relation, would give us the relation among the entropy, the mass  and the external field hairs  of black holes after matching the charge function $c(r)$ with those conserved charges in black hole solutions. In some cases, this matching might be made model-independently by using the `on-shell' scaling path in the solution space as was given in the previous section through the quasi-local conserved charge formalism~\cite{Kim:2013zha,Kim:2013cor,Hyun:2014kfa,Hyun:2014sha}.

We would like to emphasize that the charge  function depends on the gauge choice but the result from the Smarr-like relation is independent of the specific gauge choice~\cite{Ahn:2015uza}. By using the boundary condition at the horizon $r=r_{H}$ described earlier,  one can see that in Einstein gravity the charge function is directly connected with the entropy, just like the case with planar symmetry, as  
\begin{equation} \label{Hor}
c(r_{H}) = \frac{(D-2)}{16\pi G}r_H e^{A(r_{H})}f'(r_{H}) = (D-2) T_{H}{\small \text{$ {\cal  S}$} }\,.
\end{equation}

The value of the charge function $c(r)$ at the asymptotic infinity could be related to the mass and, if exist, $U(1)$ charges of the planar black holes, once again as in the case with full planar symmetry. However, contrary to the case of the full planar symmetry,  the matching of charge function $c(r\rightarrow\infty)$ with mass or $U(1)$ charges would be somewhat delicate, since $c(r)$ may contain the divergent expression as $r$ goes to infinity. This behavior of $c(r)$ can be inferred from the Smarr-like relation in Eq.~(\ref{ChargeFunc}) as follows. As is well-known, the on-shell action value  in the asymptotic AdS-geometry diverges.  Therefore, we immediately see that  the right-hand side of the Smarr-like relation in Eq.~(\ref{ChargeFunc}) would become divergent generically as $r$ goes to  infinity. By subtracting the divergent part consistently in both sides of the equality in Eq.~(\ref{ChargeFunc}), one can relate the mass and $U(1)$ charge of planar black holes with the value of charge function $c(r)$ at infinity and then, the final expression would look like the  relation among conserved quantities.  This consistent subtraction may be achieved, for instance, by the background subtraction\footnote{Though  the background subtraction method produces a physically sensible result  in many cases,   there are some cases in which the background subtraction does not give a unique answer or it cannot be used~\cite{Emparan:1999pm,Mann:2005yr}.   Our adoption of the background subtraction is just for convenience, not essential in our argument. We leave  other choices of the consistent subtraction  as our future works.} in the bulk computation, which corresponds to the specific holographic renormalization scheme. 
%
As is clear from the rewritten expression of Smarr-like relation in the form of
\begin{equation} \label{FiniteSmarr}
c(r_{H}) = (D-2) T_{H}{\small \text{$ {\cal  S}$} } = \bigg[ c(r) - \frac{1}{16\pi G} \int^{r}_{r_{H}} dr  \sum_{\omega}(\omega+1)L_{[\omega]}\Big|_{on-shell}\bigg]_{r\rightarrow\infty}\,,
 \end{equation}
one can see that the combination of the right-hand side should be finite and independent of the renormalization scheme. 
\\
\\
\underline{\it The implication of the Smarr-like relation to the dual boundary theory}

Now, we apply our Smarr-like relation to  holographic systems and reveal its implication in the thermodynamics of the dual systems. Since our holographic models are supposed to describe the homogeneous systems, their grand potential, ${\cal W}$, in the dual field theory is equivalent to the pressure: $ {\cal P} =-{\cal W}$.  In our cases, the grand potential  is given by
\begin{equation} \label{thermo rel}
{\cal W} = {\Large \text{$\epsilon$}} -\mu  Q  - T S \,,
\end{equation}
where $\mu$,  $Q$,  $T$ and $S$ denote the chemical potential, the charge, the temperature and the entropy of the dual boundary system, respectively. According to  the AdS/CFT correspondence, these   correspond to  the chemical potential $\mu$, the $U(1)$ charge ${\cal Q}$, the black hole temperature $T_H$ and the entropy ${\small \text{$ {\cal  S}$}}$ of bulk planar black holes, respectively.
The computation of the energy, ${\Large \text{$\epsilon$}}$,  of the dual boundary system typically involves non-trivial renormalization scheme with counter terms. As the counter terms depend not only on the Lagrangian but also on 
the boundary conditions of
the background fields, the determination of the counter terms is very difficult and tedious part in the computation. On the other hand, from  the AdS/CFT correspondence the energy  ${\Large \text{$\epsilon$}}$  of the dual boundary system    corresponds to the ADT mass $M$ of bulk planar black holes which can be straightforwardly computed by utilizing the simple background subtraction method.

By using the expression in Eq.~(\ref{FiniteSmarr}), we now reveal the implication of our novel Smarr-like relation  to the dual field theory.
In the context of the AdS/CFT correspondence,  the grand potential is nothing but the renormalizd on-shell action
: $T_{H}{\cal I}^{ren}_{on-shell} = {\cal W} = -{\cal P}$. Hence, one can see that
\begin{align}   \label{ren action}
T_{H}{\cal I}^{ren}_{on-shell} 
= M - \mu {\cal Q} -  \frac{1}{D-2}
 \lim_{r \to \infty}
  \bigg[ c(r) -  \frac{1}{16\pi G}\int_{r_H}^r dr  \sum_{\omega}(\omega+1)L_{[\omega]}   \bigg]_{on-shell}\,,
\end{align}
 where  $M$, $\CQ$ and $\mu$ are obtained from the bulk computation as  in~\cite{Wald:1993nt,Iyer:1994ys,Wald:1999wa,Kim:2013zha,Kim:2013cor} through the  background subtraction method. 
This explicit expression of the finite on-shell renormalized action  is another main result in this paper. Namely, it could be regarded as an efficient way
to obtain the holographic renormalized action in the dual field theory.
This bulk computation would be related to a scheme in the holographic renormalization. Nevertheless, with the AdS/CFT correspondence, we do not need to know the details of the scheme.
In particular, we do not need to bother to construct  various counter terms  which are needed to cancel the divergences from   planar symmetry breaking terms in the Lagrangian: $\sum_{\omega\neq -1} L_{[\omega]}|_{on-shell}$.

Some specific examples are presented in the next section. Our results  are checked and confirmed by using the conventional counter term method in  Appendix B. Through the given examples, one may see the effectiveness of our approach in these specific quantities in AdS/CMT models. 

%

\section{Application to holographic models}\label{application}

In this section we would like to  see the implication of our results concretely. To this purpose, we will focus on the examples in the asymptotic $AdS_{4}$ space, whose dual field theory describes three-dimensional system.

\subsection{Comparison to known results}

Let us consider a holographic model describing a dual CMT  in the form of 
\begin{align}\label{Lagrangian00}
I[g,{\cal A},\varphi_I,\psi] &= \frac{1}{16 \pi G} \int d^4 x \, \sqrt{-g}\, \bigg[ R +6 -\frac{1}{4} \CF^{\mu\nu} \CF_{\mu\nu}  -|D_\mu \psi|^2 - m_\psi^2 |\psi|^2 \\&\qquad\qquad-m^2_1 \,\mathcal U_1 - m_2^2 \,\mathcal U_2 -\frac{1}{2} \Big( 1+\frac{q_\varphi}{8\sqrt{-g}}\epsilon^{\mu\nu\rho\sigma}{\cal F}_{\mu\nu}{\cal F}_{\rho\sigma} \Big) \sum_{I=1}^{2} \partial_{\mu}  \varphi_I  \p^{\mu} \varphi_I \bigg] \,,\nn
\end{align}
where the massless scalar fields $\varphi_I$ and the $\CF\wedge \CF$-term are introduced for a momentum relaxation and a nontrivial magnetization~\cite{Seo:2015pug}, respectively. The complex scalar field $\psi$
is introduced to describe scalar condensates. 
The mass operators for the metric are given by  $\mathcal U_1= {\mathcal{K}^\mu}_\mu$ and $\mathcal U_2=({\mathcal{K}^\mu}_\mu)^2 - {\mathcal{K}^\mu}_\nu {\mathcal{K}^\nu}_\mu$, where $\mathcal{K}$ is defined by ${(\mathcal{K}^2)^\mu}_\nu \equiv g^{\mu\alpha}f_{\alpha\nu}$ and $f_{\mu\nu}$ is the reference metric given by a diagonal matrix with eigenvalues, $(0,0,1,1)$~\cite{Vegh:2013sk}.

With the metric ansatz in Eq.~(\ref{ansatz}),
we take the gauge field and the scalar fields as follows:
\begin{align}\label{ansatz0}
 \CA  &= \frac{a(r)}{r^2} dt-\half H y dx +\half H x dy\,, \qquad \varphi_{I} = ({\b x}\,\,, {\b y})\,,  \qquad \psi=  \left\{
  \begin{array}{cc}
    \psi(r)& ( H = 0 )\\
    0& (H\neq 0) \\
  \end{array}
\right.
\,,
\end{align}
where  the parameters $H$ and $\beta$ denote the external magnetic field and   the strength of  the momentum relaxation  in the dual field theory system, respectively. Note that   the phase for the complex scalar $\psi$ is taken trivial.
Taking all the ingredients into account, we arrive at the reduced action  which can be divided into the ``invariant'' sector $L_{[-1]}$ and the ``broken'' sector $L_{[\omega \neq -1]}$ under the scaling transformation as follows:
%
%
%
%
\begin{align}\label{redL}
L_{red} = &L_{[-1]}~+\sum_{\omega\neq -1}L_{[\omega]} \,,
\end{align}
where
\begin{align}\label{redL1}
&L_{[-1]} =- \frac{e^A}{r} \bigg[ 2rf'+6f -6   +r^2f|\psi'|^2 - m_\psi^2|\psi|^2   \bigg]+ \frac{ (r a'{}-2 a )^2}{2 r e^A} +\frac{q_\psi^2 a^2|\psi|^2}{re^A f} \,, \\
&L_{[-2]}= -\sqrt{2} m_1^2 \frac{e^A}{r^2},~L_{[-3]}=- \frac{e^A}{r^3} \bigg[  \b^2 + m_2^2  \bigg],~L_{[-5]}= \frac{2 q_{\varphi}\beta^{2}H(r a'-2 a)-e^A H^2 }{2r^5}\nonumber  \,.
\end{align}
%


\noindent\underline{\it{Spontaneously magnetized system}: $\left( \psi(r) = m_1=m_2 =0 \right)$}

The first class of models  depicts magnetized systems with the external field.
This case covers various AdS/CMT models and was studied in~\cite{Hartnoll:2007ih, Kim:2015wba, Blake:2015ina, Seo:2015pug}.
Note that the dual system has nontrivial magnetization even without the external field.
It is straightforward to obtain the expression of the charge  function $c(r)$ as
\begin{equation} \label{}
c(r) = \frac{ e^A }{8\pi G}\Big[ r f'-  e^{-2A} a (r a'-2a)\Big]
 - \frac{1}{8\pi G}\frac{ q_{\varphi} \beta^{2}H a }{r^4} \,.
\end{equation}
%
%
The analytic $AdS_4$ black hole solution to EOM has been obtained in~\cite{Seo:2015tug} and the relevant functions are given by
\begin{align}\label{Magnetic Solution}
f(r) &= 1 - \frac{\b^2}{2r^2} - \frac{m}{r^3} + \frac{H^2 + \mathcal{Q}^2}{4 r^4} -\frac{q_\varphi \mathcal{Q} H \b^2}{6 r^6} + \frac{q_{\varphi}^2 H^2 \b^4}{20 r^8} \,, \\
a(r) &= r^2\Big(16\pi G \mu - \frac{\mathcal{Q}}{r} + \frac{q_\varphi H \b^2}{3 r^3}\Big)\,, \qquad  A(r) =3\log r\,, \nn 
\end{align}
where the $U(1)$ charge is defined by
\[   
\mathcal{Q}\equiv -\sqrt{-g}\mathcal{F}^{rt}-\frac{q_\varphi}{4}\epsilon^{rt\mu\nu}\mathcal{F}_{\mu\nu}(\partial \varphi)^2\,.
\]
In this case, our Smarr-like relation in Eq.~(\ref{FiniteSmarr}) tells us that
\begin{equation} \label{} 
 T_{H}{\small \text{$ {\cal  S}$} }\ = \frac{3}{2}M + \frac{\beta^{2} r_H}{16\pi G}  - \mu \mathcal{Q}  + \frac{1}{16\pi G}\bigg( \frac{2 q_{\varphi} \mathcal{Q} \b^2}{3 r_H^3} - \frac{H (5 r_H^4 + 2 q_{\varphi}^2\, \b^4)}{5 r_H^5} \bigg)H\,,
\end{equation}
where the black hole mass $M=m/(8\pi G)$ is identified by using the ADT method. At this point,  we would like to show the cancellation of the divergence  between two terms in the right-hand side of the equality in Eq.~(\ref{FiniteSmarr}).  The explicit forms of divergences of both terms  are  given by
\begin{align}
c(\Lambda)|_{on-shell}&= \frac{\beta ^2 \Lambda }{8 \pi  G}+\left(3 M-2 \mu  \mathcal{Q} \right) +\mathcal{O}\left( \frac{1}{\Lambda}\right)\,,  \\
\frac{1}{16\pi G}\int_{r_H}^\Lambda dr& \sum_\omega (\omega+1) L_{[\omega]}|_{on-shell} \\
&=\frac{\beta ^2 \Lambda }{8 \pi  G}+\frac{3 H^2 \left(5 r_H^4+2 \beta ^4 q_{\varphi }^2\right)-10 \beta ^2 H \mathcal{Q} r_H^2 q_{\varphi }-15 \beta ^2 r_H^6}{120 \pi  G r_H^5} +\mathcal{O}\left( \frac{1}{\Lambda}\right)\nonumber\,.
\end{align} 
Indeed, the leading divergent terms cancel each other and thus the Smarr-like relation becomes the following form:
\begin{equation} \label{Smarr1}
M - T_{H} {\small \text{$ {\cal  S}$} } - \mu \mathcal{Q} = - \Big[\frac{M}{2} + \mathcal{M}_\beta \beta+ {\cal M} H \Big]\,,
\end{equation}
where   $\mathcal M H$ is the magnetic subtraction term and $\mathcal{M}_\beta \beta$ is the similar subtraction term from the external axion field~\cite{Seo:2015pug,Andrade:2013gsa}. The magnetization and the response to an external field $\beta$ can be read off as
\begin{align}
\CM = \frac{\beta ^2 \mathcal{Q} q_{\varphi }}{24 \pi  G r_H^3}-\frac{H \left(5 r_H^4+2 \beta ^4 q_{\varphi }^2\right)}{80 \pi  G r_H^5}\,, \qquad \mathcal M_{\beta}=\frac{\beta r_H}{16\pi G}\,.
\end{align}

In the end, by using the result in Eq.~(\ref{Smarr1}), one can get the pressure as
\begin{equation} \label{}
{\cal P} = \frac{M}{2}+ \mathcal{M}_\beta \beta+ \mathcal M H\,,
\end{equation}
which is the same expression as the one obtained by a holographic renormalization method in \cite{Kim:2015wba}. We present the brief summary of the holographic method in  Appendix B, which shows  the efficiency of our approach by  the Smarr-like relation.\newline

\noindent\underline{\it{Massive gravity models for holographic lattice}: $\left(\psi(r)=\beta=q_\varphi=0\right)$}

The second example is a  phenomenological model in which the graviton masses describe effectively a lattice structure in the dual system~\cite{Vegh:2013sk,Blake:2013owa}. This model admits an analytic solution  \cite{Vegh:2013sk}:
\begin{align}\label{massiveSol}
f(r) &= 1 -\frac{m_1^2}{2 \sqrt{2}r} - \frac{m_2^2}{2r^2} - \frac{m}{r^3} + \frac{H^2 + \mathcal{Q}^2}{4 r^4} \,,\nn\\
a(r) &=r^2( 16 \pi G\mu - \frac{\mathcal{Q}}{r}) \,, \qquad A(r)=3 \log r\,.\nn
\end{align}
By
the general procedure in the previous section,
the corresponding charge function is easily obtained as
\begin{equation} \label{massive charge}
c(r) = \frac{ e^A }{8\pi G}\Big[ r f'-  e^{-2A} a (r a'-2 a)\Big]\,.
\end{equation}
The charge function becomes divergent near the boundary as follows:
\begin{align}
c(\Lambda) = \frac{\Lambda ^2 m_1^2}{16 \sqrt{2} \pi  G}+\frac{\Lambda  m_2^2}{8 \pi  G}+\left( 3 M-2 \mu  \mathcal{Q}\right)+ \mathcal{O}\left(\frac{1}{\Lambda} \right)\,,
\end{align}
which is canceled with the other term as in the previous example.

%
%

By plugging the above solution  to Eq.~(\ref{massive charge}),
the Smarr-like relation together with our general formula in Eq.~(\ref{ren action}) leads to
\begin{equation} \label{}
{\cal P} = \frac{M}{2}  + \mathcal{M}H + \frac{m_1^2 r_H^{2}}{16\pi G \sqrt{2}} + \frac{m_2^2 r_{H}}{8\pi G}\,,
\end{equation}
where ${\mathcal M}= - \frac{H}{16\pi G r_H}$. The last two terms of the pressure can be interpreted as the contributions from the holographic lattice~\cite{Cai:2014znn}.  
In this case, the mass of the planar $AdS_{4}$ black holes is also computed as  $M = m/8\pi G$. Even though the Lagrangian parameters $m_1$ and $m_2$ are taken as transforming parameters in the on-shell parameter variation path, one can still show that the same expression of the mass $M=m/(8\pi G)$ is obtained.
Our approach does not have the ambiguity as in~\cite{Blake:2013bqa,Cao:2015cza}, which comes from  the choice of counter terms.\newline


\noindent\underline{\it{Modified holographic superconductor} : ($m_1=m_2= q_{\varphi}= H =0$)}

The final example describes the broken phase in the dual system  by turning on the nontrivial charged scalar hair, $\psi(r)$.  This covers the models for superconductor transition in \cite{Hartnoll:2008vx,Hartnoll:2008kx,Kim:2015dna,Andrade:2014xca}.
In this model  the charge function $c(r)$ is given by
\begin{equation} 
c(r) = \frac{ e^A }{8\pi G}\Big[ r f'+ r^2 f \psi'^2 -e^{-2A} a (r a'-2 a)+ q_\psi^2 e^{-2A}f^{-1} a^2\psi^2\Big]\,.
\end{equation}
%
%
The Smarr-like relation in the Eq.~(\ref{ChargeFunc}) becomes
%
%
\begin{equation}\label{SmarrSuper}
c(r) - c(r_H) =  \frac{1}{8\pi G}\int_{r_H}^{r}dr \beta^2 \frac{e^{A(r)}}{r^3}\,,
\end{equation}
where the charge function   behaves near the boundary as 
\begin{align}
c(\Lambda)=\frac{\beta ^2 \Lambda }{8 \pi  G}+3 M-2 \mu  \mathcal{Q}+ \mathcal O \left(\frac{1}{\Lambda}\right)\,.
\end{align}  
The divergence in this case is also canceled by the divergence from the right-hand side  in Eq.~(\ref{SmarrSuper}).
By using boundary expansions of the fields as the Eq.~(\ref{asymptotic}),  agreed with those in \cite{Kim:2015dna},
%
%
and using the general formula in Eq.~(\ref{ren action}),
%
%
we arrive at
\begin{equation} \label{}
\mathcal P =  \frac{M}{2} + \frac{\beta^2 r_H}{8\pi G} - \frac{1}{8\pi G}\int_{r_H}^\infty dr \beta^2 \left( \frac{e^{A(r)}}{r^3} -1 \right)\,,
\end{equation}
where the last term in the right-hand side gives pressure coming from interaction between the superfluid degrees of freedom and the impurity or a lattice effect described by the axion. This is completely matched with the on-shell action in \cite{Kim:2015dna} through the holographic renormalization.

In all the examples given in this section,  our general formula for the on-shell renormalized action in Eq.~(\ref{ren action})
gives the same expression as the one obtained from holographic renormalization.  See Appendix B for some details of the holographic renormalization approach. One may note that in our formalism we do not need
any explicit expression of counter terms in contrast to the holographic renormalization approach.


%
\subsection{On-shell action for a general class of relaxed Einstein-Maxwell-dilaton theories}

In this section, let us consider  a holographic model which describes the metal insulator transition~\cite{Donos:2012js,Donos:2014uba}. We start with the following action:
\begin{align}
I_{EMD} =
 \frac{1}{16 \pi G} \int d^4 x \, \sqrt{-g}\, \bigg[ R - V(\phi) -\frac{Z(\phi)}{4} \CF^{\mu\nu} \CF_{\mu\nu}  -\frac{1}{2}  (\partial \phi)^2 -\frac{1}{2} \sum_{I=1}^2  \Phi_I(\phi)   (\p \varphi_I)^2    \bigg] \,. 
\end{align}
We take scalar field   as $\phi=\phi(r)$ and $\varphi_I = \left( \beta_1 x ,\beta_2 y\right)$ with the same  ansatz (\ref{ansatz}) and (\ref{ansatz0}) for  the metric and gauge fields.
We assume that the geometry is an asymptotically AdS space, thus $V(\phi) \to- 6$ as $r\to \infty$. By following the same procedures given in the previous section, one can construct the reduced action  and find the charge function and the Smarr-like relation.

Finally, we find the on-shell renormalized action as follows:
\begin{align}
T_{H}{\cal I}^{ren}_{on-shell}
\nn
=&M -  \frac{1}{32\pi G}\lim_{r\to \infty}\left[e^{A(r)} \left(  2 r f'(r) + f(r) r^2 \phi'(r)^2\right) \right]_{on-shell}\\
&+ \frac{1}{32\pi G}\int_{r_H}^\infty dr \Big[e^{A}\left(\frac{2 H^2   Z(\phi)}{r^5}
+\beta_1^2  \frac{\Phi _1(\phi)}{r^3}+\beta_2^2  \frac{\Phi _2(\phi)}{r^3}\right) \Big]_{on-shell}~~.
\end{align}
As a by-product, one can easily read off the magnetization, ${\cal M}= - \int_{r_H}^\infty dr ~\frac{e^A H Z(\phi)}{16\pi G r^5}$ by employing the asymptotic behaviors of the fields  in Eq.~(\ref{asymptotic}),
which is in the complete agreement with the expression of the magnetization in~\cite{Blake:2015ina}.

\section{Conclusion}
It has been known that the reduced action, which describes some specific sector of the original action, may have an accidental symmetry which has nothing to do with the original symmetry. Interestingly, this symmetry of the reduced action may be related to the on-shell symmetry or not. In the context of the planar black holes considered in our paper, we have introduced the `scaling symmetry/transformation' and found its consequences.
By using the `scaling symmetry/transformation' of the reduced action, we have obtained the novel Smarr-like relation in a generic model in the bulk gravity. In  various  holographic models, we have checked that our results are completely consistent with known ones and showed its effectiveness. Furthermore, our results explain the origin of the observed  identity among physical quantities, which was found empirically in some specific AdS/CMT models without noticing its universal nature.

We would like to emphasize that all the results in our method  are obtained solely from the bulk computation, which does not require any boundary counter term subtraction, except the conventional background subtraction.   As is shown by the comparison  with the holographic stress tensor computation, our method gives us the  thermodynamic relation and the on-shell renormalized action in a universal way.   It would be interesting to extend to our interpretation in a more generic way by using the direct correspondence between the bulk quasi-local ADT potential/charge and boundary ADT current/charge~\cite{Hyun:2014sha}. Another interesting direction is to investigate the relation of our study to the extended Smarr-like relation through the variation of the cosmological constant~\cite{Brenna:2015pqa}.
It would also be  interesting to extend our results to more general cases, for instance, inhomogeneous systems. For future reference in this direction, we provide some formulae in Appendix A.
 \newpage

\vskip 1cm
\centerline{\large \bf Acknowledgments}
\vskip0.5cm
{We would like to thank  Dongsu Bak and Jaehoon Jeong for helpful discussions. SH was supported by the National Research Foundation of Korea(NRF) grant
with the grant number NRF-2013R1A1A2011548. This work started from a discussion in ``Year-End Workshop of Future-Oriented HEP Society'' supported by APCTP, so we appreciate it.
KK was supported by the National Research Foundation of Korea(NRF) grant
with the grant number NRF-2015R1D1A1A01058220. SY was supported by the National Research Foundation of Korea(NRF) grant
with the grant number NRF-2015R1D1A1A09057057.}

{\center \section*{Appendix}}

\renewcommand{\theequation}{A.\arabic{equation}}
  \setcounter{equation}{0}


\section*{A. Generic model in D-dimension}
In this appendix, we present a general model in $D$-dimensions, which describe a $(D-1)$-dimensional holographic condensed matter system. Explicitly, we consider the action which is taken in the form of
\begin{equation} \label{action}
I[g, \CA, \CB, \varphi^{I}, \psi ] = \frac{1}{16\pi G}\int d^{D}x\, \sqrt{-g} \Big(\CL_{g} + \CL_{\CA} + \CL_{\CB} +\CL_{\varphi} +\CL_{\psi}  \Big) +  \frac{1}{16\pi G}\int \CL_{\CF\wedge\CH} \,,
\end{equation}
where the Lagrangians  for various fields  are given by
\begin{align}
\CL_{g}  &= R - 2\Lambda - m_g^2\,\CU(g) \,, \\ \CL_{\CA} &= - \frac{\CN_\CA(\varphi)}{4} \CF_{\mu\nu}\CF^{\mu\nu} \,, \qquad \qquad ~\quad  ~\qquad \CL_{\CB} = -\frac{\CN_\CB(\f)}{2(D-2)!} \CH_{\rho_1\cdots\rho_{(D-2)}}\CH^{\rho_1\cdots\rho_{(D-2)}} \,,\nn\\
\CL_{\varphi} &= - \frac{1}{2} G_{IJ}(\varphi)\, \partial_{\mu}{\varphi}^{I} \partial^{\mu}\varphi^{J}  - V_{\varphi}(\varphi)\,, \qquad
\CL_\psi = -|\CD_{\mu}\psi|^2-V_{\psi}(|\psi|^2)\,,\nn\\
\CL_{\CF\wedge\CH}
 &= - \frac{1}{2}  q_{IJ}~ \CF^{(2)}\wedge\CH^{(D-2)}\, \partial_{\alpha}{\varphi}^{I} \partial^{\alpha}\varphi^{J} \,,\nn
\end{align}
where $\CD_{\mu}$ denotes the gauge covariant derivative defined by $\CD_{\mu} = \p_{\mu} - iq_\psi\CA_{\mu}$ and  $q_{IJ}$ are constants.
Here, $\CF = d\CA$ and $\CH = d\CB$ denote the 2-form and $(D-2)$-form field strength, respectively.
Note that the ${\cal U}$-term in ${\cal L}_{g}$ depends on the metric, not on the derivatives of the metric, and so breaks the general covariance. 

Using the ansatz for the metric in Eq.~(\ref{ansatz}) and taking the ansatz for the matter fields as
\begin{align}
\f=\f(r,{\bf x})\,,\quad\psi=\psi(r)\,,\quad\CA=r^{-(D-2)}a(r)dt\,,\quad \CB= (H x_1)dx^2 \wedge \cdots\wedge dx^{D-2}\,,
\end{align}
with a constant parameter $H$, one obtains the reduced action as
\begin{equation} \label{}
I_{red}[f,A,a,\varphi^I,\psi] = \frac{1}{16\pi G} \int dr d{\bf x}\,(L_{g} + L_{\CA} + L_{\CB} + L_{\varphi}+L_{\psi} +L_{\CF\wedge\CH} )\,,
\end{equation}
where
\begin{align}   \label{}
&L_{g} \equiv - \frac{e^A}{r} \bigg[ (D-2)\Big( (D-1) f +rf' \Big) + 2\L + m_g^2\, \CU(g) \bigg] \,,\\
&L_{\CA} \equiv \frac{\CN_A}{2r e^A} \Big( ra'-(D-2)a \Big)^2  \,,\qquad \qquad  L_{\CB} \equiv -\frac{\CN_\CB}{2r}e^A r^{-2(D-2)}H^2 \,,  \nn \\
&L_{\varphi} \equiv - \frac{e^A}{r} \bigg[  V_\f(\f) + \frac{1}{2} G_{IJ}(\varphi) f r^2 \f^{I}{}' \f^{J}{}'  \bigg] -\frac{e^A}{2r^3}G_{IJ} \partial_i \f^{I} \partial_i \f^{J} \,,  \nn \\
&L_{\psi} \equiv - \frac{e^{A}}{r} \bigg[ V_\psi(|\psi|^2)+ f |r\psi'|^2  - \frac{q_\psi^2 a^2 |\psi|^2}{e^{2A}f} \bigg]  \,,\nn \\
&L_{\CF\wedge\CH} \equiv \frac{q_{IJ} H }{2r^{D-1}}\Big( ra'-(D-2)a \Big)\Big( fr^2 \f^I{}' \f^J{}' +\frac{\partial_i\f^I \partial_i\f^J}{r^2}\Big)\,. \nn 
\end{align}
%
Note that the total derivative terms in the reduced action are irrelevant, and so are omitted in the above.
It is straightforward to obtain the charge density function $c(r, {\bf x})$ under the scaling transformation, whose explicit form is given as follows:
\begin{align}\label{}
16\pi G \,c(r,{\bf x}) &= e^A \bigg[ (D-2)rf' + G_{IJ} f r^2 \f^{I}{}'\f^{J}{}' +f  r^2 |\psi'|^2 \bigg] \\
&\qquad -\frac{\CN(\f)}{ e^A} \Big( ra'-(D-2)a\Big)(D-2)a + \frac{ 2q_\psi^2 a^2 |\psi|^2 }{  e^A f }\nn\\
&\qquad -\frac{q_{IJ}H}{2r^{D-2}}\bigg[ \Big(2ra'-(D-2)a\Big)fr^2\f^I{}'\f^J{}' + (D-2)a\Big( \frac{\partial_i \f^{I} \partial_i \f^{J}}{r^2} \Big)\bigg]\,\nn
\end{align}
where we have used  $\CE_{\Psi}=0$ in order to eliminate the scalar potentials $V_\f(\f)$ and $V_\phi(|\phi|^2)$.
In the case of $D=4$, this charge function reproduces results in the  section 3 .

\renewcommand{\theequation}{B.\arabic{equation}}
  \setcounter{equation}{0}


\section*{B. Holographic renormalization method}

In this section we summarize a method to obtain the on-shell action and the Smarr-like relation through the institutive holographic renormalization \cite{Balasubramanian:1999re,Henningson:1998gx,deHaro:2000vlm,Skenderis:2002wp}. We consider two examples in Sec.~\ref{application}. The first example admits  an analytic solution while the other has a numerical solution.

\subsection*{B.1. Spontaneous magnetized system}  
In this section, we summarize the holographic method for the spontaneous magnetized system~\cite{Kim:2015wba}. As is analyzed in Sec.~\ref{application}, this system has an analytic solution written in Eq.~(\ref{Magnetic Solution}). In this case we have to consider two kinds of boundary actions such as the Gibbons-Hawking and the counter terms obtained by the holographic renormalization procedure \cite{Henningson:1998gx,deHaro:2000vlm,Skenderis:2002wp}.  

The total renormalized action is given by $I_{ren} = I_{bulk} + I_{GH} + I_{counter}$ and  the bulk action $I_{bulk}$ is taken as Eq.~(\ref{Lagrangian00}) with $\psi(r)=m_1=m_2=0$. The other parts are given as follows\footnote{In the following, we take $16\pi G$ and AdS radius $L$ as 1 for simplicity.}:
\begin{align}
&I_{GH} = -2 \int_{r=\Lambda} d^3 x \sqrt{-\gamma}~K \\
&I_{counter} = \int_{r=\Lambda} d^3 x \sqrt{-\gamma} \left(-4 + \frac{1}{2}\sum_{J=1}^2\gamma^{ab} \partial_a \varphi_J \partial_b \varphi_J \right)~~,
\end{align}
where $K$ is the trace of the extrinsic curvature tensor $K_{\mu\nu}\equiv \nabla_{(\mu} n_{\nu)}$ with a outward normal vector $n^\mu$ along the holographic direction. And $\gamma$ is the determinant of the induced metric $\gamma_{ab}$ coming from the ADM decomposition as follows:
\begin{align}
ds^2 =N^2 dr^2 +  \gamma_{ab} \Big( dx^{a}+ N^a dr  \Big) \Big(dx^{b}+ N^b dr \Big)\,.
\end{align}
Here the coordinates ${x}^a$ denote the boundary coordinates $(t,x,y)$.
Following the decomposition, our general ansatz gives
\begin{align}
N =\frac{1}{r\sqrt{f(r)}}\,, \qquad N^a = 0\,, \qquad \gamma_{ab}=diag\left(-r^{-4}e^{2A(r)} f(r)~,~r^2~,~r^2\right)\,.
\end{align}
We have chosen $I_{counter}$  as  a popular one  and  we may choose another scheme by adding finite term to this counter term. The divergence structures of each part are given by
\begin{align}
&
I_{bulk} =\int d^3 x \left[ -2 \Lambda ^3 + \frac{-3 H^2 \left(5 r_H^4+3 \beta ^4 q_{\varphi }^2\right)+10 \beta ^2 H \mathcal{Q} r_H^2 q_{\varphi }+15 r_H^4 \left(4 r_H^4+\mathcal{Q}^2\right)}{30 r_H^5}+ \mathcal O \left( \frac{1}{\Lambda}\right) \right]\,, \nonumber\\
&I_{GH} =\int d^3 x \left[6 \Lambda ^3-2 \beta ^2 \Lambda -3 m + \mathcal O \left( \frac{1}{\Lambda}\right)\right]\,,\nonumber\\
&I_{counter}=\int d^3 x\left[-4 \Lambda ^3 +2 \beta ^2 \Lambda +2 m+ \mathcal O \left( \frac{1}{\Lambda}\right)\right]\,.
\end{align}

Now we are ready to get the on-shell action. By summing the above terms, one can obtain the Euclidean on-shell action: 
\begin{align}
T_H \mathcal{I}^{ren}_{on-shell} =\frac{3 \beta ^4 H^2 q_{\varphi }^2}{10 r_H^5}+\frac{H^2}{2 r_H}-\frac{\beta ^2 H \mathcal{Q} q_{\varphi }}{3 r_H^3}-\frac{\mathcal{Q}^2}{2 r_H}-2 r_H^3+m\,.
\end{align}
To check the Smarr-like relation, we need to compute boundary energy-momentum tensor given by the following formula: 
\begin{align}
T_{ab} =& \lim_{\Lambda \to \infty}\frac{2 \Lambda}{\sqrt{-\gamma}}\frac{\delta I_{ren}}{\delta \gamma^{ab}}\nonumber \\
&=\lim_{\Lambda \to \infty} 2 \Lambda \Big[K_{ab}-K \gamma_{ab} - 2\gamma_{ab} - \frac{1}{2}\sum_{J=1}^2 \partial_a \varphi_J \partial_b \varphi_J + \frac{1}{4}\gamma_{ab}(\partial \varphi_J)^2  \Big]_{r=\Lambda}~.
\end{align}
From this one evaluates the  energy ${\Large \text{$\epsilon$}}=T_{00}= 2m = M$. Together with the pressure expression ${\cal P} = -T_{H}{\cal I}^{ren}_{on-shell}$, we have checked that this system satisfies   the Smarr-like relation: 
\begin{align}
{\Large \text{$\epsilon$}} + \mathcal P = \mu {\cal Q} + T_HS\,.
\end{align}

\subsection*{B.2. Modified holographic superconductor}

Now let us consider a system in which only numerical solution  is available. The action $I_{bulk}$ for this case is given by Eq.~(\ref{Lagrangian00}) with $m_1=m_2=q_\varphi = H = 0$. For simplicity, we take $m^2_\psi = -2$. 
Then, this scalar field $\psi$ is dual to either $\Delta=1$ or $\Delta=2$ operator and one  may choose the counter term as follows: 
\begin{equation} \label{}
I_{counter} = \int_{r=\Lambda} d^3 x \sqrt{-\gamma} \left[-4 + \frac{1}{2}\sum_{J=1}^2\gamma^{ab} \partial_a \varphi_J \partial_b \varphi_J + \eta_1\, \psi^* \psi   + \eta_2\, n^\mu\left( \psi^* \partial_\mu \psi +c.c.  \right)  \right]\,,  
\end{equation}
where $\eta_1$ and $\eta_2$ can be chosen case by case. For instance, when the bulk complex scalar field $\psi$ is dual to a dimension $\Delta=1$ operator, we take the term proportional to  $\eta_2$  as the Gibbons-Hawking-like term. In this subsection, we will consider the $\Delta = 2$ case, which corresponds to  $\eta_2=0$ and $\eta_1 = -1$. 

Combining the $(x,x)$ component of the Einstein equation and the trace of the Einstein tensor,  the bulk action $I_{bulk}$ is computed in the  following form:
\begin{align}
I_{bulk} =& \int d^4x \sqrt{-g}\left[  -g^{tt} G_{tt}-g^{rr} G_{rr}-\frac{\beta ^2}{r^2}  \right] \nonumber\\
=& \int d^3x \int_{r_H}^\Lambda dr \left[ \left(-2 e^{A(r)} f(r)\right)' - \beta^2 \frac{e^A}{r^3} ~\right]\nonumber\\
=&\int d^3x \left[-\beta ^2 \int_{r_H}^{\Lambda }  \frac{e^{A(r)}}{r^3}   \, dr-2 \Lambda ^3+\beta ^2 \Lambda +2 m + \mathcal O \left(\frac{1}{\Lambda} \right) \right]\,.
\end{align}
And the boundary terms are given as follows :
\begin{align}
&I_{GH} =\int d^3x \left[  6 \Lambda ^3-2 \beta ^2 \Lambda -3 m + \mathcal O \left(\frac{1}{\Lambda} \right) \right]\,, \nn \\
&I_{counter} =\int d^3x \left[ -4 \Lambda ^3+2 \beta ^2 \Lambda +2 m+ \mathcal O \left(\frac{1}{\Lambda} \right)\right]\,.
\end{align}
Together with all the above expressions, we arrive at the Euclidean renormalized action as follows:
\begin{align}
T_H \mathcal{I}_{on-shell}^{ren}= -\mathcal P= - m - \beta^2 r_H  - \beta^2 \int_{r_H}^\Lambda dr \left( \frac{e^A}{r^3} - 1 \right) \,.
\end{align}

The boundary energy-momentum is given by the quasi-local tensor and the variations of the counter terms. 
\begin{align}
T_{ab} & = \lim_{\Lambda \to \infty}\frac{2 \Lambda}{\sqrt{-\gamma}}\frac{\delta I_{ren}}{\delta \gamma^{ab}}  \nn \\
&=\lim_{\Lambda \to \infty} 2 \Lambda \Big[K_{ab}-K \gamma_{ab} - 2\gamma_{ab} - \frac{1}{2}\sum_{J=1}^2 \partial_a \varphi_J \partial_b \varphi_J + \frac{1}{4}\gamma_{ab}(\partial \varphi_J)^2 - \frac{1}{2}\gamma_{ab} |\psi|^2 \Big]_{r=\Lambda}~.
\end{align}
Using this formula and the boundary expansion of the fields, we compute the holographic energy,
\begin{align}
{\Large \text{$\epsilon$}} =T_{00} = 2m\,.
\end{align}
To check the Smarr-like relation ${\Large \text{$\epsilon$}} -T_{H}{\cal S}  - \mu {\cal Q}  = - \mathcal P$, we have to solve the equation of motion by a numerical shooting method. In such a method, values and derivatives of the fields at the horizon can be regarded as the initial shooting parameters which are encoded in ${\cal S}$ and $T_{H}$. On the other hand ${\Large \text{$\epsilon$}}$, $\mu$ and ${\cal Q}$ can be read off from the boundary behaviors of the numerical solution. In addition the pressure needs a numerical integration using the solution. Therefore, our analytic derivation of the Smarr-like relation is very useful to check the numerical approach.
%


%
%

%

\end{document}